\newcommand{\lya}{\mbox{Ly$\alpha$}}
\newcommand{\halpha}{\mbox{H$\alpha$}}
\newcommand{\hbeta}{\mbox{H$\beta$}}
\newcommand{\hgamma}{\mbox{H$\gamma$}}
\newcommand{\heII}{\mbox{He~{\sc ii}}}
\newcommand{\oII}{\mbox{O~{\sc ii}}}
\newcommand{\oIII}{\mbox{O~{\sc iii}}}
\newcommand{\cIV}{\mbox{C~{\sc iv}}}
\newcommand{\neIII}{\mbox{Ne~{\sc iii}}}
\newcommand{\neV}{\mbox{Ne~{\sc v}}}
\newcommand{\siIII}{\mbox{Si~{\sc iii}}}
\newcommand{\mgII}{\mbox{Mg~{\sc ii}}}
\newcommand{\kms}{\mbox{km\,s$^{-1}$}}
\newcommand{\squarcsec}{\mbox{$\square^{\prime\prime}$}}
\newcommand{\Muv}{\mbox{$M_\mathrm{UV}$}}
\newcommand{\Lstar}{\mbox{$L^\star$}}
\newcommand{\nodata}{...}

\documentclass[a4paper,fleqn,usenatbib,twocolumn]{aastex631}
\usepackage{newtxtext,newtxmath}
\usepackage[T1]{fontenc}
\usepackage{xcolor}
\usepackage{longtable}
\usepackage{tablefootnote}
\usepackage{graphicx}	
\usepackage{amsmath}

\received{int' nu}
\accepted{men nu!}
\submitjournal{ApJ Letters}

\shorttitle{High redshift variable sources in the HUDF}
\shortauthors{M. J. Hayes et al.}

\begin{document}

\title{Glimmers in the Cosmic Dawn: A Census of the Youngest Supermassive Black Holes by Photometric Variability
\footnote{This research is based on observations made with the NASA/ESA Hubble Space Telescope obtained from the Space Telescope Science Institute, which is operated by the Association of Universities for Research in Astronomy, Inc., under NASA contract NAS 5–26555. These observations are associated with programs 1563,12498,17073.}}

\correspondingauthor{Matthew J. Hayes}
\email{matthew.hayes@astro.su.se}

\author[0000-0001-8587-218X]{Matthew J. Hayes}
\affiliation{Stockholm University, Department of Astronomy and Oskar Klein Centre for Cosmoparticle Physics, AlbaNova University Centre, SE-10691, Stockholm, Sweden.}

\author[0000-0002-3389-9142]{Jonathan C. Tan}
\affiliation{Department of Space, Earth \& Environment, Chalmers University of Technology, SE-412 96 Gothenburg, Sweden}
\affiliation{Department of Astronomy, University of Virginia, Charlottesville, VA 22904, USA}

\author[0000-0001-7782-7071]{Richard S. Ellis}
\affiliation{Department of Physics and Astronomy, University College London, Gower Street, London WC1E 6BT, UK}

\author[0000-0001-9136-3701]{Alice R. Young}
\affiliation{Stockholm University, Department of Astronomy and Oskar Klein Centre for Cosmoparticle Physics, AlbaNova University Centre, SE-10691, Stockholm, Sweden.}

\author[0000-0002-2070-9047]{Vieri Cammelli}
\affiliation{Dipartimento di Fisica, Sezione di Astronomia, Università degli Studi di Trieste, via G.B. Tiepolo 11, I-34131, Trieste, Italy}
\affiliation{INAF - Osservatorio Astronomico di Trieste, via G.B. Tiepolo 11, I-34131, Trieste, Italy}
\affiliation{Department of Space, Earth \& Environment, Chalmers University of Technology, SE-412 96 Gothenburg, Sweden}

\author[0000-0002-6260-1165]{Jasbir Singh}
\affiliation{INAF – Osservatorio Astronomico di Brera, via Brera 28, I-20121 Milano, Italy}
\affiliation{Dipartimento di Fisica, Sezione di Astronomia, Università degli Studi di Trieste, via G.B. Tiepolo 11, I-34131, Trieste, Italy}
\affiliation{Department of Space, Earth \& Environment, Chalmers University of Technology, SE-412 96 Gothenburg, Sweden}

\author[0000-0002-1025-7569]{Axel Runnholm}
\affiliation{Stockholm University, Department of Astronomy and Oskar Klein Centre for Cosmoparticle Physics, AlbaNova University Centre, SE-10691, Stockholm, Sweden.}

\author[0000-0001-5333-9970]{Aayush Saxena}
\affiliation{Department of Physics, University of Oxford, Denys Wilkinson Building, Keble Road, Oxford OX1 3RH, UK.}

\author[0000-0001-9454-4639]{Ragnhild Lunnan}
\affiliation{Stockholm University, Department of Astronomy and Oskar Klein Centre for Cosmoparticle Physics, AlbaNova University Centre, SE-10691, Stockholm, Sweden.}

\author[0000-0002-9642-7193]{Benjamin W. Keller}
\affiliation{Department of Physics and Materials Science, University of Memphis, 3720 Alumni Avenue, Memphis, TN 38152, USA}

\author[0000-0003-2083-7564]{Pierluigi Monaco}
\affiliation{Dipartimento di Fisica, Sezione di Astronomia, Università degli Studi di Trieste, via G.B. Tiepolo 11, I-34131, Trieste, Italy}
\affiliation{INAF - Osservatorio Astronomico di Trieste, via G.B. Tiepolo 11, I-34131, Trieste, Italy}
\affiliation{INFN, Sezione di Trieste, Via Valerio 2, 34127 Trieste TS, Italy}
\affiliation{IFPU, Institute for Fundamental Physics of the Universe, via Beirut 2, 34151 Trieste, Italy}

\author[0000-0001-7459-6335]{Nicolas Laporte}
\affiliation{Aix Marseille Université, CNRS, CNES, LAM (Laboratoire d’Astrophysique de Marseille),
UMR 7326, 13388 Marseille, France.}

\author[0000-0003-0470-8754]{Jens Melinder}
\affiliation{Stockholm University, Department of Astronomy and Oskar Klein Centre for Cosmoparticle Physics, AlbaNova University Centre, SE-10691, Stockholm, Sweden.}

\begin{abstract} 
We report first results from a deep near infrared campaign with the Hubble Space Telescope to obtain late-epoch images of the Hubble Ultra-Deep Field (HUDF), 10-15 years after the first epoch data were obtained. The main objectives are to search for faint active galactic nuclei (AGN) at high redshifts by virtue of their photometric variability, and measure (or constrain) the comoving number density of supermassive black holes (SMBHs), $n_{\rm SMBH}$, at early times. In this Letter we present an overview of the program and preliminary results concerning eight objects. Three variables are supernovae, two of which are apparently hostless with indeterminable redshifts, although one has previously been recorded at a $z\approx 6$ object precisely because of its transient nature. Two further objects are clear AGN at $z=2.0$ and 3.2, based on morphology and/or infrared spectroscopy from JWST. Three variable targets are identified at $z= 6- 7$, which are also likely AGN candidates. These sources provide a first measure of $n_{\rm SMBH}$ in the reionization epoch by photometric variability, which places a firm lower limit of $3\times 10^{-4}\:{\rm cMpc}^{-3}$. After accounting for variability and luminosity incompleteness, we estimate $n_{\rm SMBH}\gtrsim 8\times 10^{-3}\:{\rm cMpc}^{-3}$, which is the largest value so far reported at these redshifts. This SMBH abundance is also strikingly similar to estimates of $n_{\rm SMBH}$ in the local Universe. We discuss how these results test various theories for SMBH formation.
\end{abstract}


\section{Introduction}\label{sect:intro}

The search for Active Galactic Nuclei (AGN) in the early Universe proceeds for several reasons.  Each counted AGN means at least one supermassive black hole (SMBH) is identified, which places a lower limit constraint on the abundance of SMBHs at early times. Accurate determination of this abundance may in turn be used to test SMBH formation theories \citep[e.g.,][]{Rees.1978, Volonteri.2010, Banik.2019, Inayoshi.2020, Singh.2023}. The recovered number density, and associated SMBH seeding mechanisms, may in turn be used as an empirically motivated input to galaxy formation simulations, where AGN are fundamentally implicated in galaxy evolution, luminosity functions, quenching, and other phenomena. 

Some theoretical models \citep[e.g.,][]{ Banik.2019} predict large numbers of AGN to exist in the reionization epoch, and recent observations from JWST may (at least in part) confirm the abundance to be larger than general community-wide expectations from before 2022 \citep[e.g.,][]{Maiolino.2023agn,Harikane.2023agn,Larson.2023}. Most specifically, the families of ``Little Red Dots” at $z>4$ \citep[][]{Furtak.2023,Matthee.2023,Wang.2024,Kokorev.2024} mark an unexpected discovery. These compact objects identified at high-$z$ with a smoothly rising red SED, and often showing broad Balmer emission lines.
While it is not settled that the majority of these contain AGN, spectroscopic studies show that a substantial fraction may do so, which could imply large numbers of AGN in the reionization era. With AGN expected to exhibit high hydrogen ionizing photon production efficiencies and escape fractions, they could (once-more) become implicated in reionization scenarios. 

The challenge to this field comes from the fact that identifying AGN at luminosity regimes of typical galaxies (below \Lstar, with \Muv$\gtrsim -20$) is observationally difficult. There are numerous tracers of accretion-powered excitation in galaxies including, but not limited to: X-ray or radio continuum emission that is too bright to be produced by star formation; the presence of broad emission lines; the presence of emission lines with ionization potentials too high to be explained by stellar photoionization or shocks; various ratios of emission line fluxes; and mid-infrared colors consistent with dust heated to temperatures $\sim1000\:$~K.  Yet the heterogeneous population of AGN typically satisfies subsets of these criteria, which become observationally more challenging/less complete at different rates as surveys target higher redshifts. This leads to SMBHs probably being under-counted, with potentially large numbers going unnoticed among the ostensibly star-forming galaxy population at high-$z$. Multiple diagnostics are mandatory to place stronger constraints on the AGN abundance. 

Here we argue that the photometric variability that results from changes in the mass accretion rate of SMBHs can provide a completely independent and complementary probe of AGN. Monitoring for variability selects AGN from imaging data directly by phenomena related to the SMBH, without any biases of photometric preselection (color, luminosity, compactness, etc), and has been used successfully at intermediate redshifts in deep HST imaging \citep[][]{Cohen.2006,Pouliasis.2019,OBrien.2024}. However, the highest redshifts require monitoring at infrared wavelengths, in order to even detect galaxies: JWST would obviously be the best tool, but as such a recent observatory it will take years to build the requisite time baselines when accounting for cosmological time dilation. HST provides a head-start. WFC3 was mounted in 2009, and almost immediately began imaging some of the best-studied extragalactic fields at $\lambda=1-1.6$~\micron. The deepest of these being the Hubble Ultradeep Field (HUDF) which was imaged in 2009 and 2012 (\citealt{Bouwens.2010z8} and \citealt{Ellis.2013}, respectively) and provides the optimal field for a search for faint AGN at high-redshifts. 

We re-imaged the HUDF in 2023, to search for variable sources over a 10-15 year time window. This is critical for very high-$z$ systems where time dilation reduces this delay by a factor of $(1+z)$, to timespans of just $\sim 1$~year in the restframe. In this Letter we present an overview of our survey and the observational datasets (Section~\ref{sect:obs}) and some selected preliminary results (Section~\ref{sect:results}) that include supernovae, independently confirmed AGN at $z=2-3.5$, and newly-discovered transients whose photometric/spectroscopic redshifts are consistent with sources in the reionization era.  We provide some discussion regarding the SMBH abundance at $z>6$ in Section~\ref{sect:disc}, and the implications for SMBH seeding mechanisms. 

\section{Observations, Data and Methods}\label{sect:obs}

\subsection{Observations \& Reductions}
In order to maximize our sensitivity to high-$z$ variable AGN, we elected to target the HUDF because this field has provided samples at the highest comoving volume density. The field was first observed in the optical with ACS \citep{Beckwith.2006}, and then with the critical NIR mode with WFC3/IR in 2008-9 under GO\,11563 (UDF09; PI: Illingworth), which conducted imaging in three filters (F105W, F125W, and F160W) over 192 orbits.  In 2012 the field was imaged again under GO\,12498 (UDF12; PI: Ellis), who substantially deepened the F105W and F160W observations, and added a fourth filter, F140W, to search for Lyman break galaxies at $z\sim8$. With the singular goal of searching for photometric variability in all sources in the HUDF IR footprint, we re-imaged the field in September 2023 in the F140W filter, matching the precise centre, field orientation, and depth (30 orbits) of the GO\,12498 observation. 

\begin{deluxetable}{cccc}
\tablehead{
\colhead{Year} & \colhead{Filters} & \colhead{Orbits} & \colhead{GO\# / PI} 
}
\startdata
2008-9  & F105W & 24 & 11563 / Illingworth \\
        & F160W & 53 &  \\
\hline
2012    & F105W & 72 & 12498 / Ellis \\
        & F140W & 30 &  \\
        & F160W & 26 &  \\
\hline
2023    & F140W & 30 & 17073 / Hayes \\
\enddata
\caption{Observing epochs and times.}
\label{tab:epochs}
\end{deluxetable}

We reduced the F140W image with the \texttt{calwfc3} pipeline and \texttt{astrodrizzle} \citep{DrizzlePac} software, using the High-Level Science products (HLSP) obtained from the Mikulski Archive for Space Telescopes (MAST) as reference images. We also re-reduced the F140W image from the UDF12 campaign to verify that our methods match the depth of the HLSP image. At the same time, we produced independent reductions of the F105W and F160W images at the 2009 and 2012 epochs, to search for variable sources over the shorter, earlier time baseline. Full details of the observations and data reduction will be presented in Cammelli et al. (in prep.) and Young et al. (in prep.). This facilitates variability searches at three epochs: the span 2009-2012 is sampled by the F105W and F160W filters, while the duration 2012-2023 is sampled by the F140W filter only (see Table~\ref{tab:epochs} for details).

\subsection{The Search for Variable Sources}

We searched for variables in each matched pair of filters using two techniques: comparison of the nuclear/central photometry of galaxies at different epochs, and the detection of residual sources in pair-subtracted images. We avoid regions toward the edge of the image where the dithering pattern has led to excess noise, and consider only a central pointing of the HUDF covering 123\arcsec$\times$139\arcsec. The photometric comparison follows a similar method to that of \citet{OBrien.2024}: for each of the six of WFC3/IR images we identify sources using \texttt{Source Extractor} \citep{Sextractor}. In contrast to performing `global' galaxy photometry (e.g. with Kron-like or moment centered apertures as \texttt{Source Extractor} would do) we seek to obtain photometry centered upon the brightest, unresolved sources within each galaxy. We therefore rely only upon the brightest pixel coordinates reported by \texttt{Source Extractor}, and the background images it produces as part of its background subtraction procedure.  We first use \texttt{astropy}'s \texttt{photutils} package to produce precisely re-centered coordinates in the immediate vicinity of the \{\texttt{XPEAK\_IMAGE},\texttt{YPEAK\_IMAGE}\} for each source.  At this coordinate we then uses \texttt{photutils} to perform aperture photometry in 4 pixel diameter (0\farcs 26) apertures with a local background subtraction to remove local (non-compact) galaxy light. However, this would not include Poissonian uncertainties from the sky background, which we calculate from the background level images stored by \texttt{Source Extractor} (stored with \texttt{CHECKIMAGE} flags) using the same determined effective gains, and add in quadrature to the error per source using the local photometry.  We verified this approach produces values comparable to \texttt{Source Extractor}'s \texttt{MAG\_APER} measurements, but with more precise centroiding inside extended sources.

For each of the six images we use for detection, we also perform photometry in identical apertures in the other image obtained in the same filter, similar to `dual image mode' in \texttt{Source Extractor}.  Thus for each image we obtain photometry of each galaxy nucleus in detection and comparison bands, working in both directions. We contrast these local aperture magnitudes in each of the images with respect to the photometric uncertainty of each source.  However we must also account for the artificially underestimated uncertainties that result from signal being correlated between adjacent pixels during the drizzling process \citep[e.g.][]{Casertano2000, Fruchter2002}.  We examine the global distribution on the $\Delta m$-vs-$m$ diagram (Cammelli et al in prep.), where $\Delta m=m_\mathrm{det} - m_\mathrm{test}$; here $m_\mathrm{det}$ is the magnitude in the image used for source detection, and $m_\mathrm{test}$ is the magnitude in the image used to test for photometric variability.

\begin{figure*}[htb!]
\noindent
\begin{center}
\includegraphics[width=0.99\linewidth]{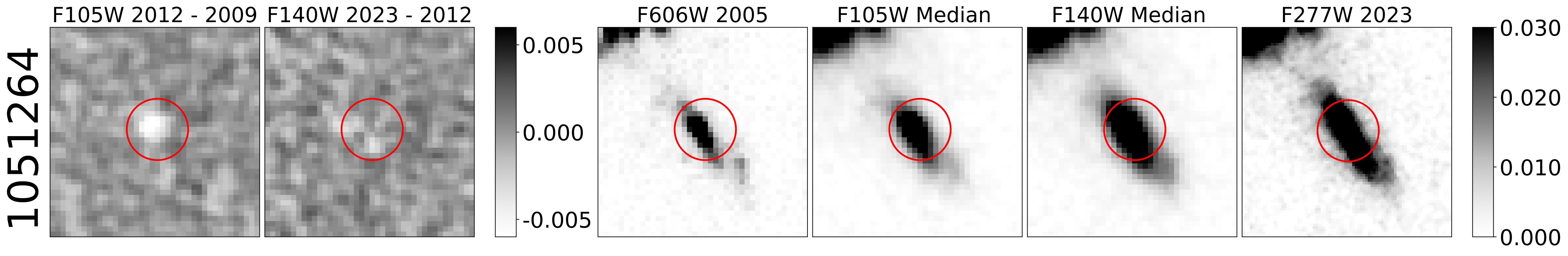}\\
\includegraphics[width=0.99\linewidth]{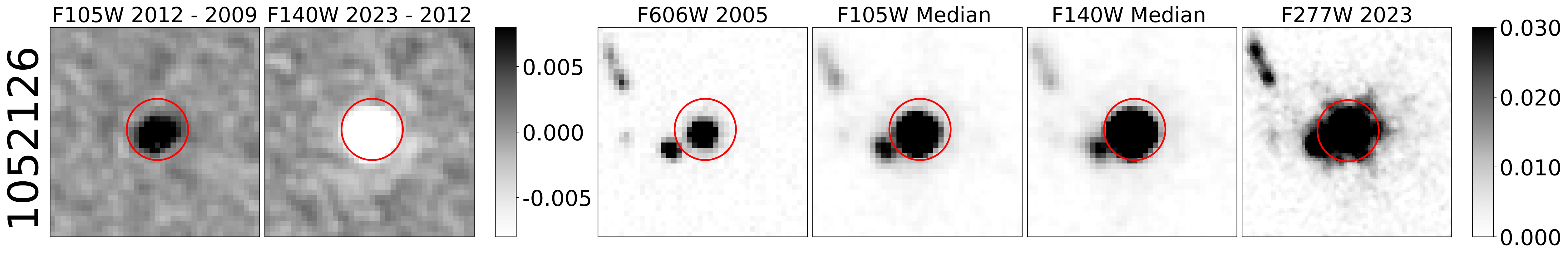}
\end{center}
\caption{Photometric variability of 1051264 at $z=2$ ({\it upper panels}) and 1052126 at $z=3.2$ ({\it lower panels}) sources. We show difference images of F105W and F140W to the left, where the earlier epoch is subtracted from the later.  The colorbar is symmetric around 0, so black objects get brighter with time and white objects have become fainter.  Panels to the right show the ACS/F606W image taken in 2005, averages of the F105W and F140W images over both epochs at which they were observed, and the JWST F277W image. Intensity images are calibrated in MJy/Sr with colorbars shown to the side. Cutouts are $2.4\times 2.4$~\squarcsec.}
\label{fig:z3phot}
\end{figure*}

\begin{deluxetable*}{ccccccccc}[htb!]
\tablehead{
\colhead{ID} & \colhead{RA} & \colhead{Dec} & \colhead{mag 2009} & \colhead{mag 2012}& \colhead{mag 2012}& \colhead{mag 2023} & \colhead{Redshift} & \colhead{Source} \\
\colhead{} & \colhead{J2000/Gaia} & \colhead{J2000/Gaia} & \colhead{F105W/F160W} & \colhead{F105W/F160W}& \colhead{F140W} & \colhead{F140W} & \colhead{} & \colhead{}
}
\startdata
1051264 & $53.180819$ & $-27.787328$ & $26.006 \pm 0.013$ & $26.332 \pm 0.013$ & $24.890 \pm 0.015$ & $24.909 \pm 0.016$ & 1.95 & NIRISS \\
1052126 & $53.178480$ & $-27.784036$ & $24.493 \pm 0.006$ & $24.378 \pm 0.003$ & $24.092 \pm 0.004$ & $24.596 \pm 0.006$ & $3.190457 \pm 0.000038$ & MUSE \\
\enddata
\caption{Properties of two example intermediate redshift AGN detected by variability.}
\label{tab:z3}
\end{deluxetable*}

We first correct a $\simeq 0.01$ magnitude systematic offset that is identified at all magnitudes, probably because of slightly imperfect zeropoints being reported to images obtained many years apart.  We then proceed by \emph{assuming that the overwhelming majority of sources in the field are not variable}.  We compute the one-sided standard deviation of the $\Delta m$ distribution in bins of 0.5~mag, where we include only sources that are \emph{brighter} in $m_\mathrm{test}$ compared to $m_\mathrm{det}$ (hence the matched-aperture photometry in both directions).  We contrast the resulting standard deviations with the propagated uncertainty on the magnitude difference derived from our photometric catalogs: $\sigma_m = (\sigma_\mathrm{det}^2 +\sigma_\mathrm{test}^2)^{1/2}$. We then compute the total scale factor that scales the combined uncertainty reported by \texttt{astropy/photutils} to reproduce the global $\Delta m$ distribution.  This ensures that, for example, 31.7\% of our sources have $\Delta m/\sigma_m \ge 1$, 4.5\% of sources have $\Delta m/\sigma_m \ge 2$, etc.  See \citep{OBrien.2024} for a more detailed discussion.

To be identified as a variable source, an object must be detected at $\Delta m/\sigma_m \ge 2$ in two or more filters, or $\Delta m/\sigma_m \ge 3$ in one filter.  Should those two filters be the F105W and F160W that were taken at the same epochs, we further demand that $\Delta m$ has the same sign (i.e. that the source is consistently inferred to be  brightening or fading over 2009-2012). If one of the filters is F140W, we enforce no sign constraints on $\Delta m$, allowing for a source to brighten then fade, or vice versa.  We have verified that for the number of sources in the images, these selections result in a negligible suspected level of contamination from non-varying sources (Cammelli et al, in prep) at magnitudes as faint as 29. For the matched sources we compute their `total significance of variability' as the quadrature addition of $\Delta m/\sigma_m$ in two bands (or three if the source is variable in all filters). 

In addition we produced pair-subtracted images for each filter to search for variable sources that may have been missed by photometric comparison, and to verify the compact nature of variables. We first smoothed each image with a very small Gaussian kernel of 0.5 pixels, and directly subtracted images obtained in the same filters.  This produced six `difference images'.  We ran \texttt{Source Extractor} on the difference images after recomputing the effective gains for each image, as the subtraction increases the sky noise and mimics a shallower integration time.  Each source was inspected by eye and spurious artefacts were removed. We retain sources that have 5-$\sigma$ significance as variable.

\begin{figure}[htb!]
\noindent
\begin{center}
\includegraphics[width=0.99\linewidth]{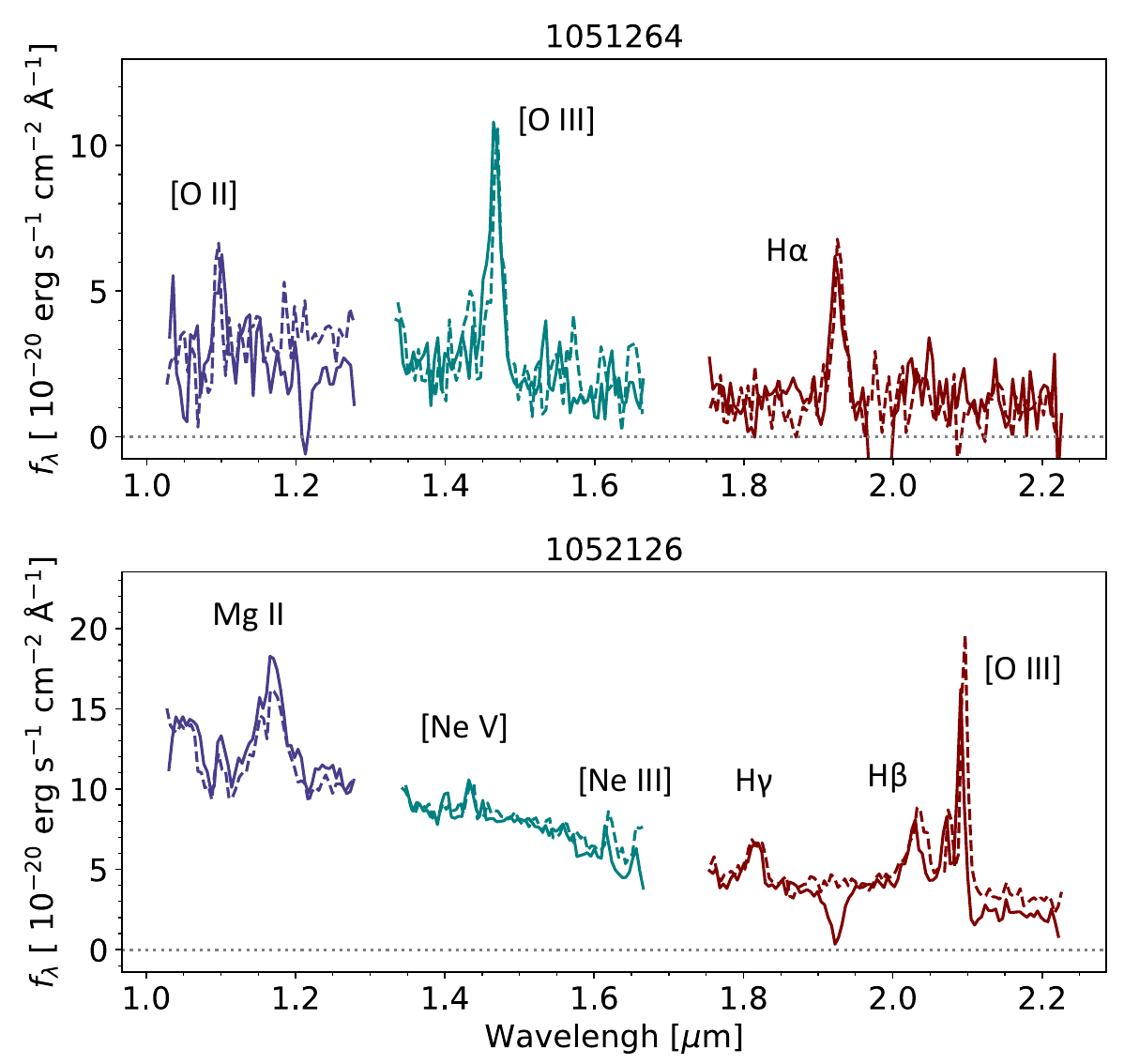}
\end{center}
\caption{JWST/NIRISS spectra of the sources shown in Figure~\ref{fig:z3phot}. The three different wavelength regions show the F115W, F150W, and F200W filters, while the two spectra per wavelength range show extractions when the spectra are dispersed in the direction of rows or columns.}
\label{fig:z3spec}
\end{figure}

\subsection{Redshift determination}
We cross-correlate each of these sources against positions in known redshift catalogs, including catalogs from: VLT/MUSE \citep[spectroscopic; ][]{Bacon.2023}; the JADES GTO program using both NIRSpec for spectroscopic redshifts \citep{Bunker.2023} and NIRCam+HST for photometric redshifts \citep{Rieke.2023}, and the Ultraviolet UDF photometric catalogs \citep{Rafelski.2015}. We also cross correlate with known AGN in the GOODS-S field, compiled by \citep{Lyu.2022}.  Further, we extract our own photometry from all HST filters (11 bands from WFC3 and ACS; \citealt{Beckwith.2006,Teplitz.2013,Bouwens.2010z8,Ellis.2013,Koekemoer.2013}), JWST in 9 broadband filters from JADES \citep{Rieke.2023} and five intermediate band filters from the JEMS program \citep{Williams.2023}.  Finally, we have reduced the deep slitless spectroscopy obtained from JWST/NIRISS \citep{Pirzkal.2023} under the NGDEEP campaign (GO\,2079; PI: Finkelstein) and extracted spectra at the location of all variable sources, recentering on sources detected in NIRISS direct imaging. See Cammelli et al. (in prep.) for the redshift distribution of the variable sources, and Young et al (in prep.) for an analysis of their morphologies and photometric inferences. 

\section{Selected variable sources in the HUDF}\label{sect:results}

In total we can identify 71 sources with photometric variability greater than $3\sigma$ over all epochs and filters. These have redshifts in the range 0--7 but are concentrated at lower-$z$.  We present preliminary results in this Letter, with the focus upon a subset of interesting variable objects.  These include two confirmed AGN at $z=2-3.2$, which demonstrate the variability method in finding convincing AGN at redshifts where they can be independently confirmed.  The HUDF is arguably the best studied region of the extragalactic sky, and has been targeted across the electromagnetic spectrum: independent AGN confirmation methods include deep optical and NIR spectroscopy from VLT/MUSE and HST/grism, the deepest X-ray imaging from Chandra, radio continuum imaging, and deep Spitzer imaging to measure MIR colors (see Section~\ref{sect:disc:method} for more details). We also show results for three transients that are most likely supernovae, of which one in particular appears at $z\simeq 6$, but is in fact an interloper at lower redshift.  We finally present our confident variable sources with redshifts in the range $z>5-7$.  

\subsection{Variable AGN at intermediate redshifts}\label{sect:results:midz}

To demonstrate the utility of photometric variability as an AGN probe, we draw attention to objects 1051264 and 1052126 in Figure~\ref{fig:z3phot} and Table~\ref{tab:z3}. 

\paragraph{Object 1051264}  This source lies in the center of an edge-on disk galaxy. Image subtraction (Figure~\ref{fig:z3phot}) shows the object is point-like.  It decreased in brightness by 0.3 magnitudes between 2009 and 2012 at very high significance (Table~\ref{tab:z3}), although the magnitude of this variation small --  ($0.08 \pm 0.025$~mag) by global (Kron) photometry because of dilution by the host galaxy. Its absolute magnitude is $M_\mathrm{UV}=-20.3$, suggesting an AGN origin.  Variability is confirmed in both F105W and F160W images but no variability was detected in 2012--2023 in the F140W image. The source is obviously point-like after the host is subtracted, and is located at the nucleus of the galaxy suggesting probable AGN activity. 

The source is identified by \citet{Rafelski.2015} and has a photometric redshift of 1.887. In Figure~\ref{fig:z3spec} we present the NIR spectra obtained using the F115W, F150W, and F200W filters of NIRISS. Emission lines of [\oII], [\oIII], and \halpha\ are unambiguous, which securely anchors the redshift to $z=1.95$, but there are no obvious signatures of AGN activity in this low-resolution spectrum.  

\paragraph{Object 1052126}   This object has a spectroscopic redshift of 3.1905 from VLT/MUSE. It is identified as a strong emitter of \lya\, \heII1640, \cIV1550, and \siIII1883, as well as other high-ionization lines \citep{Bacon.2023} and is clearly an AGN.  It is comparably bright, with $m_\mathrm{F140}=24.1$ and has varied by $-0.53$ and $+0.50$ magnitudes from 2009--2012 and 2012--2023, respectively (brightening then dimming). The flux in the variable component alone amounts to an absolute magnitude of $-21.3$, corresponding approximately to the luminosity of an \Lstar\ galaxy.  

The complex of [\oIII]$\lambda\lambda$4959,5007\AA\ and \hbeta\ is unambiguous in the NIRISS spectrum (Figure~\ref{fig:z3spec}) at about 2.1~\micron, anchoring the redshift to $z=3.2$ which is fully consistent with VLT/MUSE.  \hgamma\ emission is also clear in the F200W filter.  The F150W filter shows an obvious line of [\neIII]$\lambda 3869$\AA, but more interestingly there is a weak line at $\lambda_\mathrm{obs}=14400$~A.  This is identified as the [\neV]$\lambda 3426$\AA\ line, which requires a source of photons with $h\nu>97$~eV or dense coronal gas with temperature $T>10^5$~K. This can only be produced by an AGN. A strong \mgII\ feature is also identified in the F115W filter.  \hbeta\ is broader than the [\oIII] lines that are observed in the same filter, with FWHM of $\simeq 4600$ and $\simeq 1400$~\kms, respectively, where [\oIII] is consistent with instrumental broadening, but \hbeta\ is not.  \hgamma\ is of course weaker, but appears to be as broad as \hbeta\, and \mgII\ shows a consistent FWHM.  Most forbidden lines are too weak to provide reliable measurements of their velocity dispersion, but from the [\oIII] lines alone we see the familiar situation where permitted lines probe gas in the broad line regions (BLR), but BLR densities are sufficiently high that forbidden lines are collisionally deexcited, and [\oIII] emission probes only the more extended narrow line emitting region.

\subsection{Supernova candidates}

\begin{figure*}[htb!]
\noindent
\begin{center}
\includegraphics[width=0.99\linewidth]{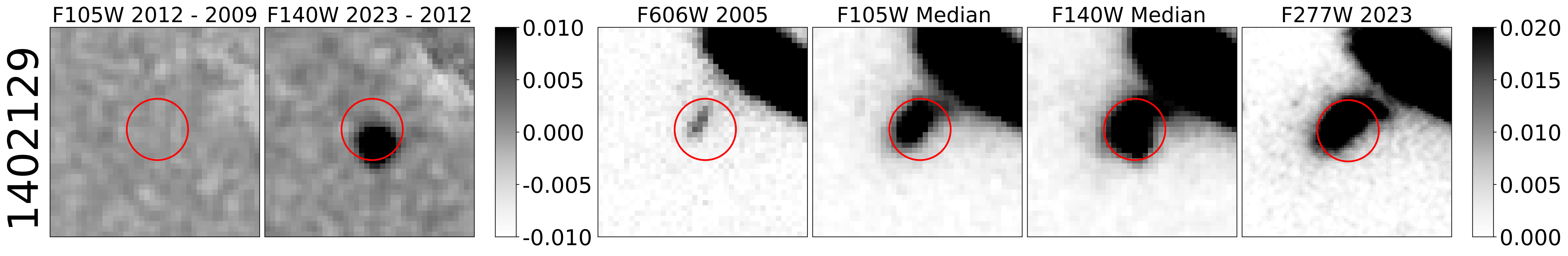}\\
\includegraphics[width=0.99\linewidth]{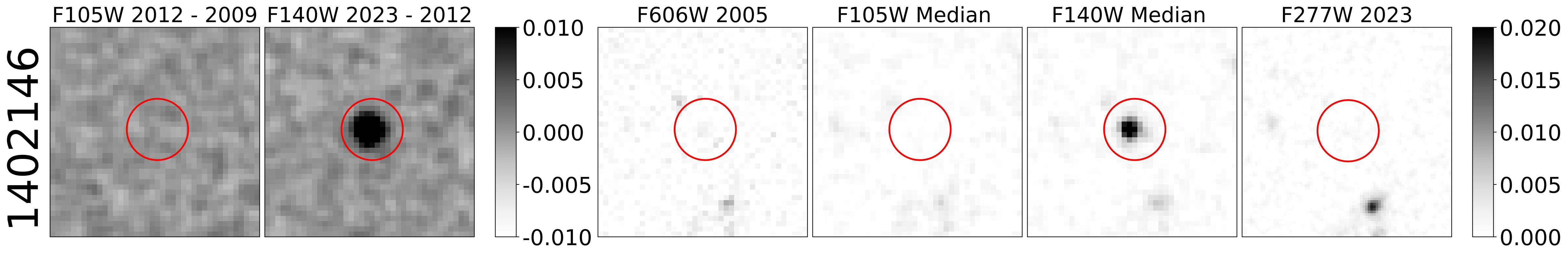}\\
\includegraphics[width=0.99\linewidth]{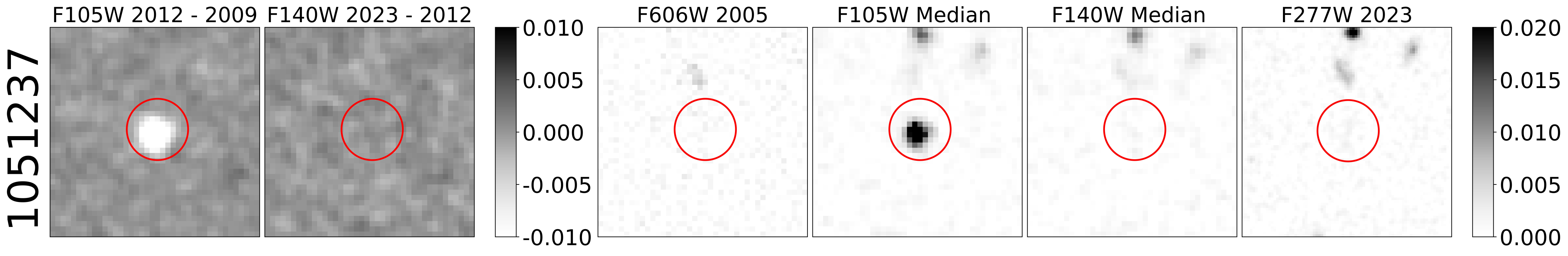}
\end{center}
\caption{Same as Figure~\ref{fig:z3phot} for the probable supernovae.}
\label{fig:snphot}
\end{figure*}

We show images of the three SN candidates in Figure~\ref{fig:snphot} and present general properties in Table~\ref{tab:sn}. 

\begin{deluxetable*}{ccccccccc}[htb!]
\tablehead{
\colhead{ID} & \colhead{RA} & \colhead{Dec} & \colhead{mag 2009} & \colhead{mag 2012}& \colhead{mag 2012}& \colhead{mag 2023} & \colhead{Redshift} & \colhead{Source} \\
\colhead{} & \colhead{J2000/Gaia} & \colhead{J2000/Gaia} & \colhead{F105W/F160W} & \colhead{F105W/F160W}& \colhead{F140W} & \colhead{F140W} & \colhead{} & \colhead{}
}
\startdata
1402129 & $53.154885$ & $-27.793421$ & \nodata            & \nodata        & $28.235 \pm 0.042$ & $26.575 \pm 0.01$ & $1.07_{-0.02}^{+0.04}$ & JADES phot \\
1402146 & $53.134812$ & $-27.788965$ & \nodata            & \nodata        & $\gtrsim 30.0$     & $27.234 \pm 0.02$ & \nodata               & \nodata \\
1051237 & $53.143873$ & $-27.793327$ & $26.927 \pm 0.017$ & $\gtrsim 30.1$ & \nodata            & \nodata           & 6.86                  & R15 \\
\enddata
\caption{Same as Table~\ref{tab:z3}, but for the SN candidate sources.}
\label{tab:sn}
\end{deluxetable*}

\paragraph{Object 1402129} This is a clear stellar transient source, exploding near the edge of a disk galaxy and is clearly not associated with nuclear activity.  {It is identified in our later epoch of F140W imaging and has increased in brightness by almost 2 mags -- a 2012 magnitude is reported in Table~\ref{tab:sn}, but this corresponds to the host galaxy in a matched aperture.  Despite being relatively bright, the host galaxy has only a redshift reported from JADES photometry, which confirms the galaxy at $z\simeq 1.07$.  With an absolute magnitude of $-16.9$ in the restframe $R$-band, the supernova could have been of any type. Since this redshift would provide a factor of only $\simeq2$ in cosmological time dilation, we expect the SN to have been relatively recent: the slowly evolving SLSNe would typically have faded by 4 magnitudes in 1 year, with ordinary type Ia and CCSNe fading faster still.

\paragraph{1402146} This is a remarkable transient source that was identified in our second epoch F140W imaging, and has exploded between 2012 and 2023.  No host galaxy is apparent in the HST/IR imaging, but a faint source is just visible in the ACS imaging.  Of our 15 visits in the observing program, one failed because of guide-star acquisition and was repeated 1 month after the remainder of the observations.  1402146 was visible also in the repeated visit and had not moved, suggesting that it lies at cosmological distances and is not a Solar System object.  The source in the ACS image is not detected in redder bands, and is so faint it does not enter any photometric redshift catalogs, so the distance to this source cannot be determined. 

\paragraph{1051237} This object appears very similar to 1402146, except it had become fainter between the 2008 and 2012 images in the F105W and F160W filters, vanishing from 26.9 mag (both filters) to undetected.  This source is intriguing, since it has a photometric redshift in the \citet{Rafelski.2015} catalog of 6.84, and appears to be a SN mistaken for a galaxy.  1051237 is visible in all the NIR filters of the HUDF09 campaign, and, while it is not visible in data from HUDF12, still appears in deep stacks that combine the two epochs.  When the ACS optical data were obtained in 2005, the source had not yet exploded, but it had subsequently faded by the time the UVUDF \citep{Teplitz.2013} imaging was obtained.  Consequently, in HST imaging only, the source is visible throughout the NIR but not in the UV/optical. In multiband photometry it therefore appears as a $z$-band dropout and is assigned a best photometric redshift of $\sim 7$.  The two galactic sources that lie about 1\arcsec\ to the north in the HST images have redshifts of 6, but other objects with redshifts in the range 0.2 to 1.1 lie within 2\arcsec. 

\begin{figure*}[htb!]
\noindent
\begin{center}
\includegraphics[width=0.99\linewidth]{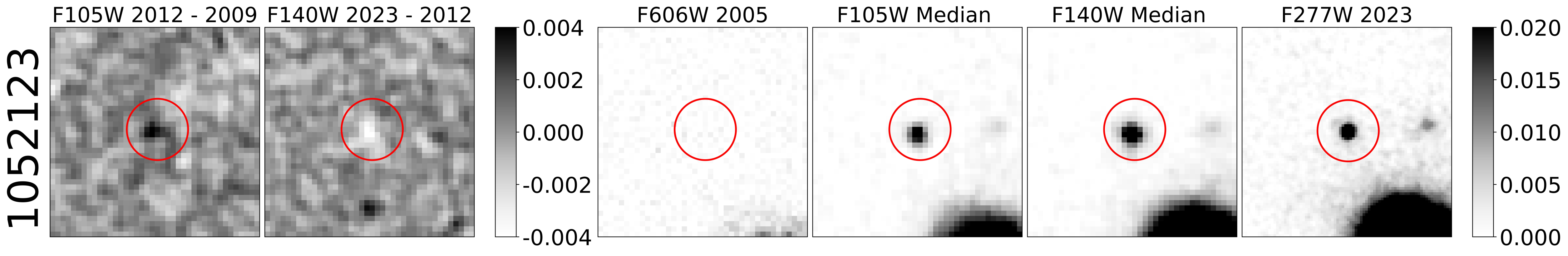}
\includegraphics[width=0.99\linewidth]{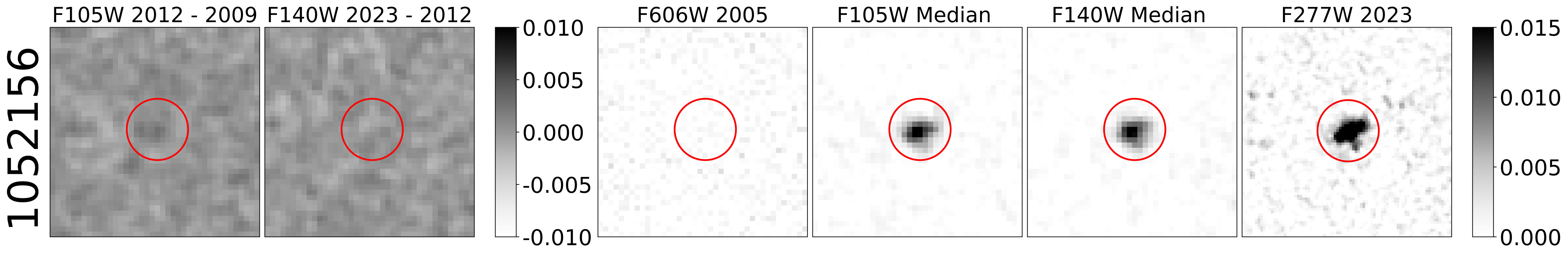}
\includegraphics[width=0.99\linewidth]{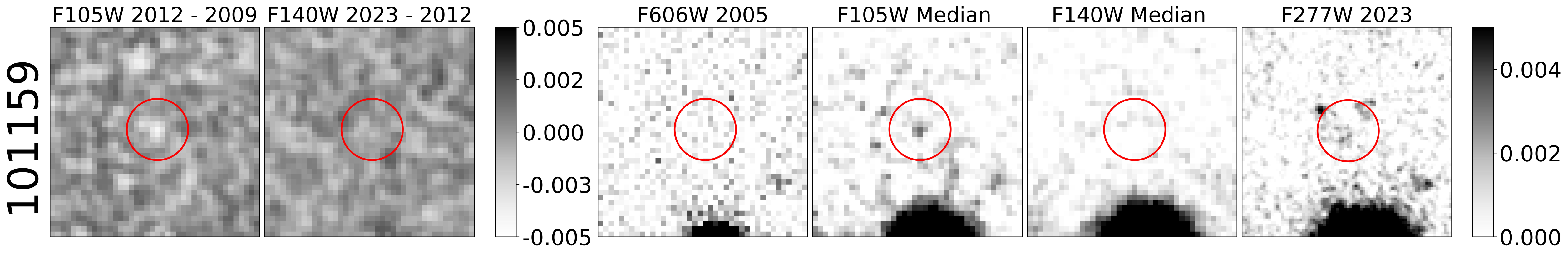}
\end{center}
\caption{Same as Figure~\ref{fig:z3phot} for the $z=6-7$ sources.}
\label{fig:z7phot}
\end{figure*}

\begin{deluxetable*}{ccccccccc}
\tablehead{
\colhead{ID} & \colhead{RA} & \colhead{Dec} & \colhead{mag 2009} & \colhead{mag 2012}& \colhead{mag 2012}& \colhead{mag 2023} & \colhead{Redshift} & \colhead{Source} \\
\colhead{} & \colhead{J2000/Gaia} & \colhead{J2000/Gaia} & \colhead{F105W/F160W} & \colhead{F105W/F160W}& \colhead{F140W} & \colhead{F140W} & \colhead{} & \colhead{}
}
\startdata
1052123  & $53.161915$ & $-27.786990$ & $28.308 \pm 0.070$ & $27.996 \pm 0.028$ & $27.595 \pm 0.024$ & $27.764 \pm 0.029$ & $6.74^{+0.073}_{-0.038}$ & JADES phot \\
1052156  & $53.154051$ & $-27.766003$ & $28.391 \pm 0.067$ & $28.181 \pm 0.034$ & $\gtrsim 29.9$     & $\gtrsim 29.9$     & $6.234$                  & JADES spec \\
101159   & $53.160519$ & $-27.785931$ & $28.961 \pm 0.366$ & $\gtrsim 30.1$ & $\gtrsim 29.9$     & $\gtrsim 29.9$     & $6.54^{+2.45}_{-2.59}$   & JADES phot \\
\enddata
\caption{General properties of the $z>6$ variable sources.}
\label{tab:z7}
\end{deluxetable*}
\subsection{Three variable sources in the reionization era}

After cross-correlating the coordinates of our variable objects with existing redshift catalogs (Section~\ref{sect:obs}), we identify three sources with significance greater than $3\sigma$ and best redshifts greater than 6.  These are, naturally, much fainter than the sources identified at mid-$z$, above.  They will be studied in detail in upcoming papers (Cammelli et al; Young et al; both in preparation) and here we present the three sources at $z>6$ that are detected at better than $3 \sigma$ in Figure~\ref{fig:z7phot} and Table~\ref{tab:z7}.  We also extracted NIRISS spectra from the NGDEEP survey, but found no convincing detections of either continuum light or emission lines in any of the three. Given the faintness of these targets, and insensitivity of low-resolution spectroscopy to weak emission lines (expected in the UV), this is not surprising. 

\paragraph{Object 1052123}  This is our highest significance variable source at $z>6$.  It is detected as variable in all three photometric bands and is visible in the pair-subtracted images of Figure~\ref{fig:z7phot}. Its light curve is shown in Figure~\ref{fig:lightcurve}.  It is point-like, and brightens between 2009 and 2012 (both F105W and F160W) by about 0.3 magnitudes, with a very similar change in brightness detected at both wavelengths implying a constant color of the variable source.   Interestingly its F105W--F140W color is 0.42 mag, while the F140W-F160W color is 0.02 mag (both measured in 2012 epoch data). This is probably due to a \lya\ break at 9400\AA, which produces a flux deficit in F105W leading to a photometric redshift of $z\simeq 6.75$, while the continuum slope is flat at redder wavelengths $(\beta = -1.88 \pm 0.05)$. The object then fades again to 2023, decreasing in brightness by 0.17 mag (4.5-sigma significance).  

The source is assigned a photometric redshift of 6.44 by \citet{Rafelski.2015} and 6.74 in the JADES redshift catalog, and we lean toward the higher value because the spectral break must lie within the F105W filter. Assuming this redshift, and that the entire variable component can be attributed to an AGN, it would have an equivalent absolute magnitude of $-19.3$ AB. This is much less luminous than the variable AGN recovered at mid-$z$ (Section~\ref{sect:results:midz}), which have $M_\mathrm{UV}=-20.3$ and $-21.9$.

The whole duration of the light curve in Figure~\ref{fig:lightcurve} is 15 years, which corresponds to 2 years in the restframe, at filters probing $\lambda_{rest} \simeq 1300-2000$\AA. These timescales exceed what is seen even for superluminous supernovae (SLSNe), which are the slowest-evolving and most UV-bright SN class known \citep{Lunnan.2018,Angus.2019,Perley.2020}. In particular, while the peak UV luminosity is well within reach for a SLSN, generally the UV fades away on a approximately months timescale due to cooling temperatures as the supernova expands \citep{Yan.2017}. Similarly, while pair-instability supernovae (PISNe) are predicted to have very slow light curves due to their extremely high ejecta masses, they are also predicted to have red spectra thanks to the copious amounts of iron-peak elements produced in the explosion, which blankets the UV flux \citep{Kasen.2011,Dessart.2013}. We therefore favor an AGN over an extreme supernova interpretation.

\begin{figure}[htb!]
\noindent
\begin{center}
\includegraphics[width=0.9\linewidth]{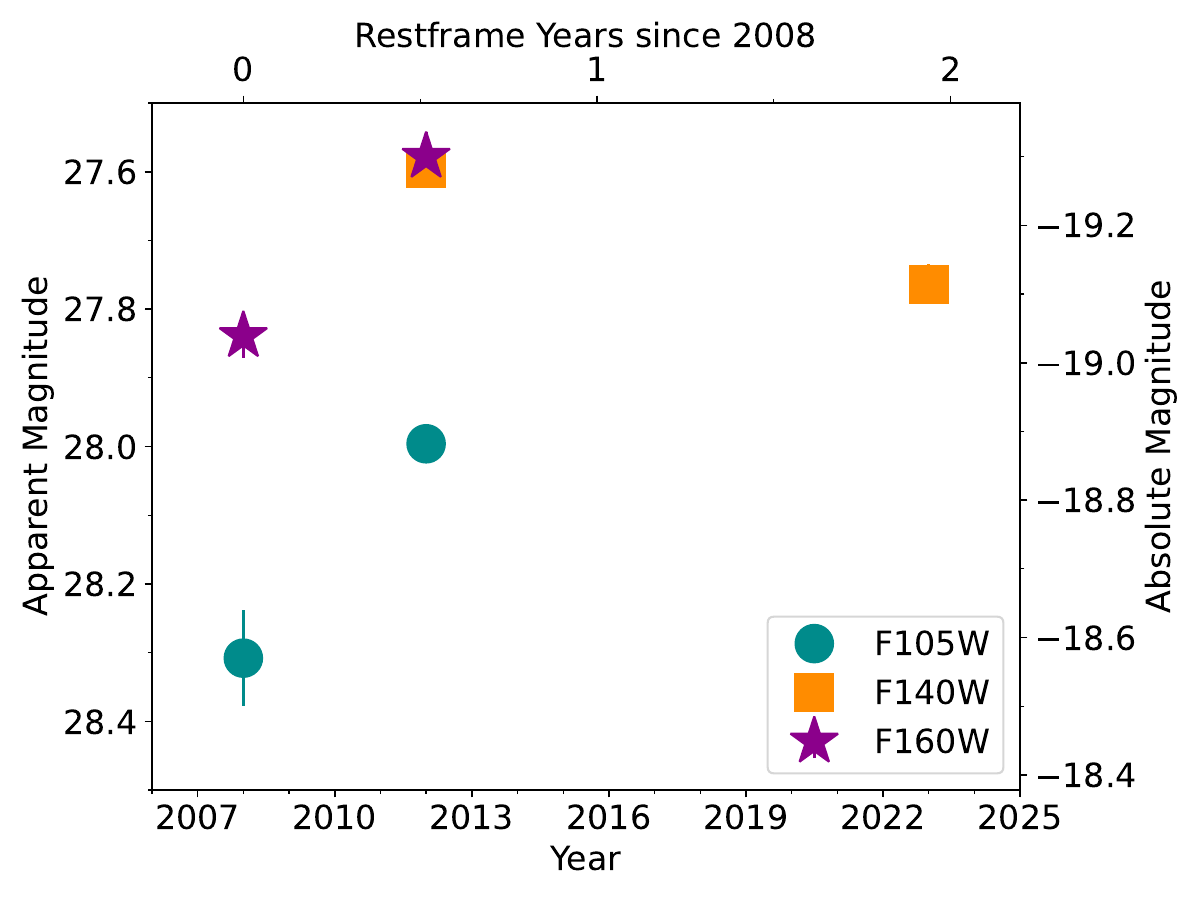}
\end{center}
\caption{Light curve for object 1052123.  Lower x axis shows the year of observation, while upper x axis shows the time since 2008 accounting for $(1+z)$.  Left axis shows the observed apparent magnitude, while the right axis shows the absolute magnitude for $z=6.74$.  Filters are shown in color.}
\label{fig:lightcurve}
\end{figure}    

\paragraph{Object 1052156} This object is identified as variable in the F105W and F160W filters (2009--2012), but is not found to be variable in the later epoch.  The variability is clear in the pair-subtracted images, and if all the variability can be attributed to a single new source, its equivalent magnitude would be $-18.5$. This magnitude is also consistent with the object being a SN, but given the change in brightness of 0.2 mags in five year, seems implausible (using the same argumentation as for 1052123).  The combined significance across the two filters is only $3.5\sigma$ in this case, but even at $\approx 28$ mag there is still not an enormous parent sample of sources, and we do not expect to find one with magnitude difference that is $3.5\sigma$ outside its distribution from pure chance.  It is therefore probable that this source is a genuine variable.  This source has a spectroscopic redshift of 6.23 \citep{Bunker.2023}, as well as various photometric redshift estimates consistent with this. 

\paragraph{Object 101159} The final object, 101159, is a remarkable variable source.  It is clearly identified by the image subtraction method as a point-like variable source that was visible in 2009, but not in 2012 or 2023.  It is not visible in the ACS optical imaging obtained in 2006, implying a possible high redshift.  It is clearly detected in the NIRCam/F277W image from JADES, which is the deepest in the JADES survey. The source likely comprises three diffuse clumps within the 1\arcsec\ region centered on the coordinates of the transient source, each of which is particularly faint, with magnitudes around $\simeq 29$ in the F277W filter.  The southern region is coincident with the location of the transient, which has no entry in the JADES catalogs. JADES reports a best fit photometric redshift for the north-western source of $z=6.54$, although the 1-sigma range is $3.84-8.99$.  The north-eastern source has a photometric redshift of 5.86, consistent with the object nearest our transient source.  If genuinely at $z>6$, the morphology of these sources is unlike most galaxies typically identified at high redshifts, where source identification is biased towards compact galaxies.  This source has unfortunately decreased in brightness since 2008.  The source could be a SLSN or an AGN, but in either case it outshines the host galaxy, which is only detectable by JWST at later epochs in redder images.  We favor the AGN scenario for similar reasons to Object 1052123: the probability of a SLSN explosion in our volume is $\sim 10^{-4}$, and these transients typically exhibit red colors and are hard to detect in the rest UV. We have verified that the source is visible in a deep stack of all the JADES imaging at NIR wavelengths, and invisible in a stack of the ACS optical filters again suggesting the object lies at high-$z$.

\paragraph{A note on interloping sources}  We initially identified another high-significance $(7\sigma)$ variable at ${\alpha,\delta}={53.1828095, -27.7760521}$, with $m_\mathrm{F140W}=28.1$. This object has a photometric redshift of 6.74 in the JADES catalog, but is assigned $z_\mathrm{phot}\simeq 1.7$ by \citet{Rafelski.2015} and is clearly detected in ACS/F606W implying a low-$z$ solution must be correct. 

\section{Discussion}\label{sect:disc}

\subsection{Variability as a method to find AGN and estimate $n_{\rm SMBH}$} \label{sect:disc:method}

We have identified several variable sources in the Hubble UltraDeep Field, using photometric monitoring at three epochs: 2008, 2012, and 2023.  In forthcoming papers (Cammelli et al., in prep; Young et al., in prep) we will present many more variable sources in the field at redshifts between 0 and 7.  Here we have presented preliminary results for eight interesting targets, two of which are clear AGN at $z\simeq2-3$, and three are probable supernovae that are also likely to have exploded at intermediate redshifts.  We find three variable sources at redshifts greater than 6.  Based upon their absolute magnitudes, which are measured at a rest wavelength of $\simeq 1300$\AA, we believe these variable sources to be AGN. 

We argue that variability searches provide a valuable and efficient technique for finding AGN in deep imaging surveys.  While only a fraction of AGN are detectable by variability, the advantage of monitoring is that it provides a complete, 100\% multiplexed survey of everything in the field with no pre-selection.  Furthermore, there is no additional challenge related to the different depths attainable for AGN diagnostics, as is the case for radio or X-ray surveys: if an object can be observed in imaging, it can be tested for variability.

Luminous quasars have been known at higher redshifts for some time \citep{Fan.2006,Mortlock.2011,Banados.2018}, which are all $\gtrsim4$ magnitudes more luminous than our sources, and potentially higher redshift AGN have also been reported \citep{Maiolino.2023agn,Harikane.2023agn,Larson.2023} at luminosities closer to \Lstar.  However, as our survey examined only the volume of the HUDF, and finds three probable AGN at $6<z<7$, this yields a higher co-moving number density of AGN than previously reported. 

Our survey samples a volume of $\simeq10,500$ comoving Mpc$^3$ between redshifts 6 and 7, leading to a number density of  $n_{\rm SMBH}\sim 2.9\times 10^{-4}$~cMpc$^{-3}$. However, only a fraction of AGN in a given luminosity range are expected to be detected by variability. \citet{Lyu.2022} have recently published a sample of confirmed AGN in the GOODS-S field, within which the HUDF is entirely contained.  31 of their AGN fall in HUDF footprint that we target, which have been identified by combinations of X-ray brightness, UV to mid-IR SED properties, optical spectral features, mid-IR colors, radio-loudness and continuum shape, and also variability. We recover 8 of these 31 AGN by photometric variability, which have redshifts between 0.6 and 3.2 (Object 1052162). \citet{Lyu.2022} identified these AGN by by MIR colors (4), X-ray luminosities (7), radio loudness (7) or optical spectroscopy (1). Some are obviously identified by more than one tracer.  Importantly, none of our 71 prime candidates have been reported previously by photometric variability: this includes the HST imaging used in \citet{Lyu.2022} who identified three variable AGN in the HUDF area, and \citet{Cohen.2006} who reported 45 further sources. We note that all this imaging is shallower, obtained at shorter wavelengths, and over shorter time baselines than our HUDF re-imaging campaigns, and should not necessarily match NIR long-baseline studies. None of our $z>6$ AGN are in the Lyu et al catalogs, almost certainly because they are too faint to be independently verified by the above techniques at current observational depths. See Cammelli et al., (2024 in prep.) for a more detailed discussion.  From the ratio 8/31 we derive a crude estimate of this variability completeness correction factor of 3.85. After such a completeness correction is applied, $n_{\rm SMBH}\sim 1.1\times 10^{-3}$~cMpc$^{-3}$.

Next, it should be noted that our survey is only sensitive to AGN above a minimum luminosity of $M_\mathrm{UV}\simeq-18.6$, so this estimate is still a lower limit on the true value of $n_{\rm SMBH}$. \citet{Harikane.2023agn} have detected AGN via the presence of broad emission lines and estimated $n_{\rm SMBH}$ from $z\sim 4$ to 7 by extrapolating down an assumed luminosity function to $M_\mathrm{UV}\simeq-17$. For a fair comparison of our results to those of \citet{Harikane.2023agn} we carry out the same method and estimate a luminosity function (assuming \Lstar$=-20$ and $\alpha=-2$) correction factor of $\sim7$, which boosts $n_{\rm SMBH}\sim 7.7\times 10^{-3}$~cMpc$^{-3}$.  Note, to the extent that not all AGN are detected down to $M_\mathrm{UV}\simeq-17$, this should still be regarded as a lower limit. In fact, at these redshifts an empirical estimate of $n_\mathrm{SMBH}$ can only be efficiently made by counting AGN, and will include only SMBHs in a rapidly accreting state.  The true comoving density would also need a final correction for the fraction of SMBHs that are currently dormant. The fact that we have no observational basis on which to estimate this final correction leads us to present all these densities as lower limits.

Our estimate of $n_{\rm SMBH}$ at $z=6-7$ in the HUDF via variability and those of \citet{Harikane.2023agn} via presence of broad spectral lines from $z=4-7$ are shown in Figure~\ref{fig:nSMBH}. Here we also show a data point at $z=0$, i.e., $n_{\rm SMBH}(z=0) \simeq 4.6\times 10^{-3}\:{\rm cMpc^{-3}}$ with an uncertainty ranging from $10^{-3}-10^{-2}\:{\rm cMpc^{-3}}$ \citep{Banik.2019}. This is calculated by assuming that each galaxy with luminosity greater than $L_{\rm min} = 0.33 \Lstar$ hosts a SMBH, with the error bar around this point assuming a range of $L_{\rm min}$ from 0.1 to 1.0~$\Lstar$ \citep[see also][for a similar estimated value of $8.79\times 10^{-3}\:{\rm cMpc}^{-3}$]{Vika.2009}. We see that our estimate of $n_{\rm SMBH}$ in the HUDF is a factor several higher than that of \citet{Harikane.2023agn} at the equivalent redshifts, but very similar to their estimate at $z\sim4$ and the local estimate at $z=0$. Figure~\ref{fig:nSMBH} also shows some predictions of theoretical models, which are discussed in the next section.

\begin{figure*}[htb!]
\noindent
\begin{center}
\includegraphics[width=0.9\linewidth]{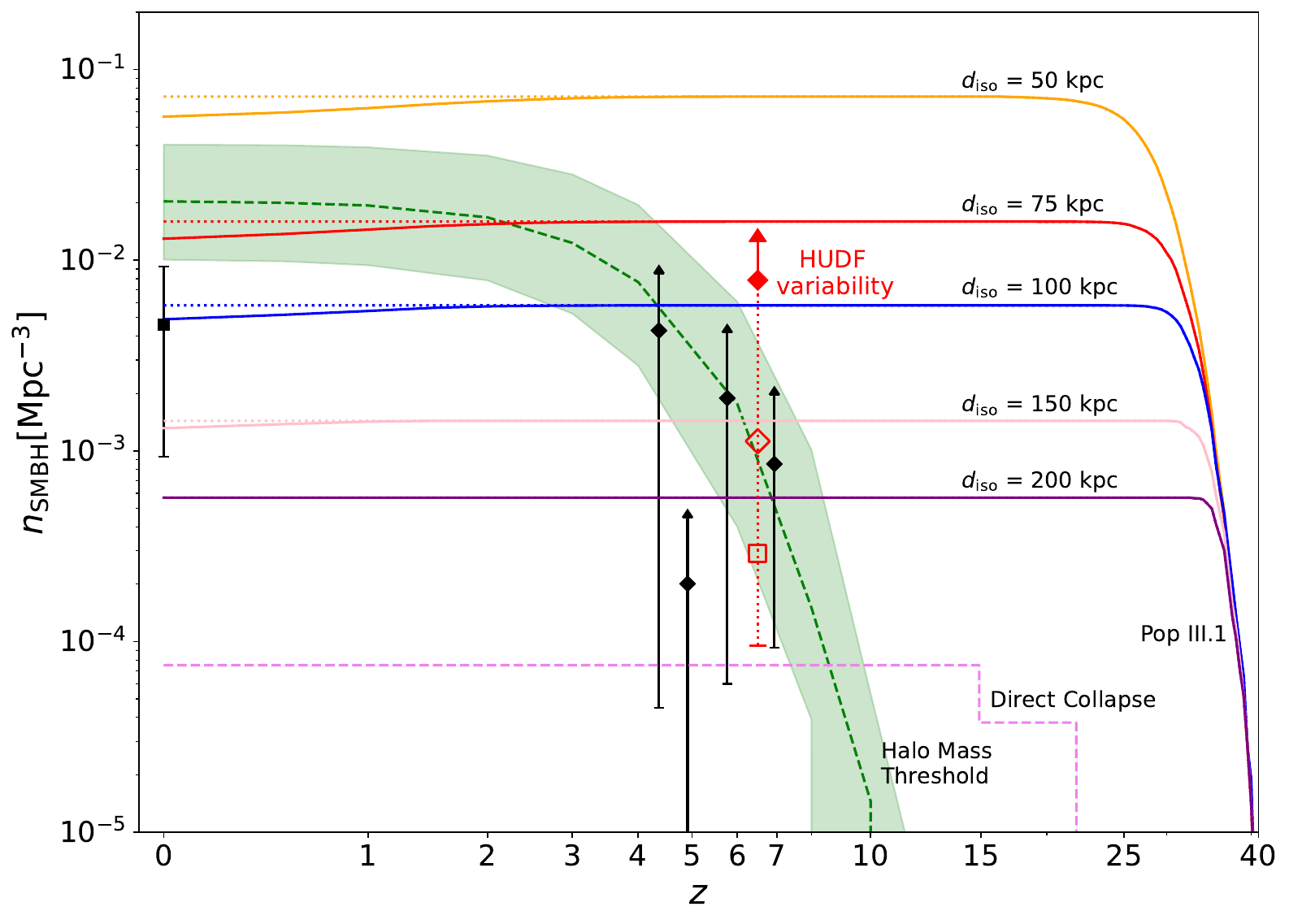}
\end{center}
\caption{Cosmic evolution of SMBH abundance. Comoving number density of SMBHs, $n_{\rm SMBH}$, is plotted versus redshift, $z$. The observational constraint derived by counting $z=6-7$ AGN in the HUDF variability survey is shown by the red open square, and the lower bound shows the Poisson uncertainty. The red open diamond corrects for variability incompleteness and the red solid diamond for luminosity incompleteness (see text). Previous observational estimates at $z=0$ from local galaxies \citep{Banik.2019} and at $z=4-7$ from broad emission line sources \citep{Harikane.2023agn} are shown with the black square and diamonds, respectively. For the $z\sim 4-7$ sources we are unable to make any correction for the fraction of non-accreting SMBHs, and leave all points as lower limits. Various Pop III.1 formation models \citep{Banik.2019,Singh.2023} with isolation distance parameters, $d_{\rm iso}$, from $50$ to $200\:$kpc (proper distance) are shown by the colored solid lines. At low redshifts these decrease compared to the maximum value attained (dotted lines) due to mergers. The green dashed line shows SMBH seeding based on a halo mass threshold (HMT) above a mass of $7.1\times10^{10}\:M_\odot$ \citep{Vogelsberger.2014}, with the shaded region illustrating a factor of two variation in this mass scale. The magenta dashed line shows results from a model of SMBH formation via ``Direct Collapse'' \citep{Chon.2016}.}
\label{fig:nSMBH}
\end{figure*}

\subsection{Implications for SMBH formation theories} 

Our estimate of $n_{\rm SMBH}(z=6-7)\gtrsim 8\times10^{-3}\:{\rm cMpc^{-3}}$ has implications for SMBH seeding schemes and formation theories. Many implementations of SMBH seeding in cosmological simulations use simple threshold conditions. For example, a ``halo mass threshold'' (HMT) model was used in the Illustris and Illustris-TNG simulations \citep[e.g.,][]{Vogelsberger.2014}, which seeded SMBHs in halos reaching a mass of $7.1\times10^{10}\:M_\odot$. The abundance of such massive halos and associated SMBHs is shown as a function of redshift in Fig.~\ref{fig:nSMBH}. By $z\simeq 6.5$, this model has $n_{\rm SMBH}\sim 10^{-3}\:{\rm cMpc}^{-3}$, but with its value rising rapidly as the universe evolves to lower redshifts. By $z=0$, it has reached $n_{\rm SMBH}\simeq 2\times 10^{-2}\:{\rm cMpc}^{-3}$, which is marginally consistent with observational estimates $n_{\rm SMBH}(z=0) \sim 10^{-3}-10^{-2}\:{\rm cMpc^{-3}}$ \citep{Vika.2009,Banik.2019}. However, note if there is a significant population of SMBHs that have been ejected from their host galaxies and/or have very low luminosities, then $n_{\rm SMBH}(z=0)$ may have been underestimated.

``Direct Collapse'' is a physical model of SMBH formation from metal-free, UV-irradiated, atomically-cooled (i.e., relatively massive) halos, which has been proposed as a way to form massive seeds early in the universe \citep[e.g.,][]{Bromm.2003}. However, simulations of this mechanism have struggled to produce sufficient numbers of SMBHs compared to the known $z=0$ population. For example, \citet{Chon.2016} found $n_{\rm SMBH}\sim 10^{-4}\:{\rm cMpc}^{-3}$, while \citet{Wise.2019} estimated a global $n_{\rm SMBH}\sim 10^{-7}-10^{-6}\:{\rm cMpc}^{-3}$.

The ``Pop III.1'' SMBH formation model \citep{Banik.2019,Singh.2023} invokes the physical mechanism of dark matter annihilation changing the structure of primordial protostars \citep{Spolyar.2008,Natarajan.2009,Rindler-Daller.2015} allowing efficient accretion of the baryonic content of their minihalos to become supermassive stars with masses of $\sim 10^5\:M_\odot$, which then collapse to SMBHs. However, only the first minihalos to form in each local region of the universe, i.e., being pristine and undisturbed by external feedback, are Pop III.1 sources and undergo this evolution. The vast majority of minihalos are Pop III.2, i.e., irradiated by UV and/or shocked by SN blastwaves leading to enhanced HD cooling and formation of relatively low-mass stars with $\sim 10\:M_\odot$ \citep{Johnson.2006,Greif.2006}. \citet{Banik.2019} and \citet{Singh.2023} have predicted the evolution of the comoving number density of SMBHs forming from Pop III.1 seeds.  The main parameter of the Pop III.1 model is the ``isolation distance" ($d_\mathrm{iso}$), defined as the minimum separation distance a Pop III.1 source needs to have from already formed sources, with this expected to be set by the physical mechanism of radiative feedback.  To reproduce the known population of low-$z$ SMBHs with $n_{\rm SMBH}(z=0) \sim 5\times 10^{-3}\:{\rm cMpc^{-3}}$ requires $d_{\rm iso}\simeq 100\:$kpc (proper distance), i.e., $\sim 3\:$cMpc at the typical epoch of formation at $z\sim 30$. These models predict that $n_{\rm SMBH}$ is approximately constant with redshift from $z\sim25$ down to $z=0$, with only modest decreases due to mergers (see Fig.~\ref{fig:nSMBH}).

The HUDF variability estimate of $n_{\rm SMBH}(z=6-7)\gtrsim 8\times10^{-3}\:{\rm cMpc^{-3}}$ places constraints on $d_{\rm iso}$ to be $\lesssim 100\:$kpc. For HMT models it requires the halo mass threshold for SMBH formation to be $\lesssim 3\times 10^{10}\:M_\odot$. However, such models begin to have more severe tension by having significantly greater abundance than the $z=0$ estimate of $n_{\rm SMBH}$. Finally, the HUDF variability estimate of $n_{\rm SMBH}$ is about $100\times$ greater than the direct collapse prediction of \citet{Chon.2016} and at least $10^4\times$ greater than that of \citet{Wise.2019}.

Another test of the Pop III.1 model that can be made from our detected AGN is to examine the minimum separation distance between AGN. Note, the size of the field of view of the HUDF at $z\sim 6$ is about 5.5~cMpc. The closest pair of high-$z$ AGN are sources 101159 and 1052123. The former has a larger uncertainty in its redshift. If these sources were at the same redshift of $z\simeq 6.7$, then their plane of sky separation corresponds to about 0.25~cMpc. Thus, we see that the Pop III.1 model with $d_{\rm iso}\simeq 100\:$kpc (proper), corresponding to comoving separation of about 3~cMpc (and for negligible relative motion by $z\sim 7$), implies that there must be a redshift difference of $\Delta z \gtrsim 0.01$ between these AGN. Future spectroscopic measurements of the redshifts of these sources can test this prediction. Similarly, another prediction of the Pop III.1 model is that all these high-$z$ AGN are powered by single SMBHs, with binary AGN only emerging at lower redshifts (Singh et al., in prep.).
 
\subsection{Future prospects for variability searches}

It is clear from the above analysis that larger samples of AGN at high-$z$ are required to reduce the statistical (i.e., Poisson) and systematic uncertainties in the estimates of $n_{\rm SMBH}$. JWST is required to push to detection of fainter AGN via variability, requiring an investment similar to that outlined in \citet{JWSTDDTreport}. However, it will take years to establish monitoring campaigns that span significant time baselines, especially when accounting for the factor $(1+z)$ in cosmological time dilation.  In contrast, HST's legacy of deep near infrared imaging already stretches back $\simeq 15$~years, providing excellent baseline for monitoring.  The deeper IR regions of the CANDELS fields \citep{Grogin.2011,Koekemoer.2011} of strong gravitational lensing clusters \citet{Lotz.2017} would be the most appropriate next targets to carry out high-$z$ AGN variability searches to build more stringent constraints on theoretical models of SMBH formation.

\facilities{HST (WFC3), JWST (NIRISS)}

\section*{Acknowledgements}
We thank Yuichi Harikane, Marta Volonteri and Naoki Yoshida for helpful discussions, and the constructive and thoughtful  report from the anonymous referee that has improved the content of the manuscript. M.H. acknowledges the support of the Swedish Research Council, Vetenskapsr{\aa}det and is Fellow of the Knut and Alice Wallenberg Foundation. J.C.T. acknowledges support from ERC Advanced Grant 788829 (MSTAR). A.Y. is funded by the Swedish National Space Agency, SNSA.  R.~L. is supported by the European Research Council (ERC) under the European Union’s Horizon Europe research and innovation programme (grant agreement No. 10104229 - TransPIre). We thank staff in The Doors public house for service that germinated the ideas that led to this project. 
All the {\it HST} data used in this paper can be found in MAST: \dataset[10.17909/7s5v-gz68]{http://dx.doi.org/10.17909/7s5v-gz68}.

\bibliographystyle{aasjournal}

\clearpage

\end{document}